\documentclass{aa}  
\pdfoutput=1
\usepackage{graphicx}
\usepackage{txfonts}
\begin{document}

   \title{Detection of the secondary eclipse of WASP-10b in the Ks-band \thanks{Based on observations collected at the Calar Alto Observatory, Almer\'ia, Spain.}}

   \author{Patricia Cruz \inst{1}, David Barrado \inst{1}, Jorge Lillo-Box \inst{1},
   		Marcos Diaz \inst{2}, Jayne Birkby \inst{3}, Mercedes L\'opez-Morales \inst{4}, 
   		Simon Hodgkin \inst{5}, Jonathan J. Fortney \inst{6}
          }

   \institute{Depto. de Astrof\'isica, Centro de Astrobiolog\'ia (INTA-CSIC), ESAC campus, P.O. Box 78, E-28691, Villanueva de la Ca\~nada, Spain\\
              \email{pcruz@cab.inta-csic.es}
         \and
             Instituto de Astronomia, Geof\'isica e Ci\^encias Atmosf\'ericas, Universidade de S\~ao Paulo (IAG/USP), S\~ao Paulo, Brazil
         \and
             Leiden Observatory, Leiden University, Niels Bohrweg 2, 2333 CA Leiden, The Netherlands
         \and
             Harvard-Smithsonian Center For Astrophysics, 60 Garden Street, Cambridge, MA 02138, USA
         \and
             Institute of Astronomy, University of Cambridge, Madingley Road, Cambridge, CB3 0HA, UK
         \and
             Department of Astronomy and Astrophysics, University of California, 1156 High Street, Santa Cruz, CA 95064, USA
             }

   \date{Received XXXXX; accepted XXXXX}

 
  \abstract
   {WASP-10b, a non-inflated hot Jupiter, was discovered around a K-dwarf in a near circular orbit ($\sim $$0.06$). Since its discovery in 2009, different published parameters for this system have led to a discussion about the size, density, and eccentricity of this exoplanet.}
   {In order to test the hypothesis of a circular orbit for WASP-10b, we have observed its secondary eclipse in the Ks-band, where the contribution of planetary light is high enough to be detected from the ground.}
   {Observations were performed with the OMEGA2000 instrument at the 3.5-meter telescope at Calar Alto (Almer\'ia, Spain), in staring mode during 5.4 continuous hours, with the telescope defocused, monitoring the target during the expected secondary eclipse. A relative light curve was generated and corrected from systematic effects, using the Principal Component Analysis (PCA) technique. The final light curve was fitted using a transit model to find the eclipse depth and a possible phase shift.}
   {The best model obtained from the Markov Chain Monte Carlo analysis resulted in an eclipse depth of $\Delta F$ of $0.137\%^{+0.013\%}_{-0.019\%}$ and a phase offset of $\Delta \phi $ of $-0.0028^{+0.0005}_{-0.0004}$. The eclipse phase offset derived from our modeling has systematic errors that were not taken into account and should not be considered as evidence of an eccentric orbit. The offset in phase obtained leads to a value for $|e\cos{\omega}|$ of $0.0044$. The derived eccentricity is too small to be of any significance.}
   {}

\authorrunning{P. Cruz et al.}
\titlerunning{Detection of the secondary eclipse of WASP-10b in the Ks-band}

   \maketitle
%

\section{Introduction}

Since its discovery by \cite{Christian09}, the exoplanet WASP-10b seemed to be an interesting object as a close-orbiting non-inflated hot Jupiter (with a density of $\rho_{p}$$\sim $$1.43$ $\rho_{J}$) orbiting a K-dwarf with an orbital period of 3.09 days and eccentricity of $\sim $$0.06$. This scenario changed when updated stellar parameters were published revealing a higher density for WASP-10b of $\sim $$3.11$ $\rho_{J}$ (Johnson et al. 2009), leading to a discussion about the real size of this exoplanet. \cite{Christian09}, \cite{Dittmann10}, and \cite{Krejcova10} have found a radius of $1.22$-$1.28$ $R_{J}$, larger than the one published by \cite{Johnson09} ($1.08\pm 0.02$ $R_{J}$).

\cite{Maciejewski11a} suggested the presence of a third body in the WASP-10 system, which would perturb the orbital motion of WASP-10b. Based on observations of eight transits, these authors have reported Transit Timing Variations (TTVs) that could be explained by a second planet with a mass of $0.1$ $M_{J}$ and an orbital period of $5.23$ days. With high-precision photometric data, \cite{Maciejewski11b} detected signatures of stellar activity, confirming previous evidence of activity in WASP-10b (Smith et al. 2009). Taking the activity into consideration in their analysis, Maciejewski and collaborators supported the results of \cite{Johnson09} by finding a smaller planetary radius of $1.03$ $R_{J}$. Spots reduce the effective stellar disk area and can lead to an overestimation of the transit depth and, therefore, the planetary radius, which could help explaining the different results obtained so far for WASP-10b (Maciejewski et al. 2011b, Barros et al. 2013, and references therein).

These authors have also questioned the orbital eccentricity of WASP-10b, and argued that it might have been overestimated because of stellar variability, favoring a circular orbit (Maciejewski et al. 2011a,b). Later, \cite{Husnoo12} did not find conclusive evidence of an eccentricity detection after reanalyzing the radial velocity measurements by \cite{Christian09}, using a Markov Chain Monte Carlo (MCMC) analysis. They supported the idea of a circular orbit, implying that correlated noise, stellar activity, or additional companions in the system could have caused an incorrect estimation for the eccentricity.

In an attempt to confirm the existence of another companion for the WASP-10 system, \cite{Barros13} have gathered eight extra high-precision transits and analyzed them in combination with the 22 previouly published transit light curves. For their analysis, Barros and collaborators have assumed a circular orbit and they have concluded that the observations are not accurate enough to confirm the presence of another planet. Alternatively, they have suggested that the observed TTVs might have been induced by stellar activity (for more details, see Barros et al. 2013).

Another way to confirm or exclude the possibility of a circular orbit for WASP-10b is by observing its secondary eclipse, which has not been detected so far for this system. The atmospheric emission properties of this planet are still unknown.

We present the first result of the Calar Alto Secondary Eclipse study (The CASE Study): the observation of a secondary eclipse of WASP-10 and, hence, the detection of its thermal emission in the Ks-band, where the contribution of planetary light is high enough to be detected from the ground.


\section{Observations and data reduction}\label{obsred}

We observed WASP-10 ($K$=$9.983$) on 2011 August $23$, under photometric conditions, when a secondary eclipse would occur assuming circular orbit\footnote{The timming of this secondary eclipse was predicted with the help of the Exoplanet Transit Database, ETD, which is maintained by Variable Star Section of Czech Astronomical Society - for more information, see http://var2.astro.cz/ETD/index.php.}. We used the Ks-band filter (at 2.14 $\mu $m) of the OMEGA2000 instrument, which is a near-infrared wide field camera, equiped with a 2k x 2k HAWAII-2 detector, mounted on the 3.5-meter telescope at the Calar Alto Observatory (CAHA) in southern Spain, with a field of view of 15.4 x 15.4 arcmin and a plate scale of 0.45 arcsec pix$^{-1}$. 
The telescope was strongly defocused, resulting in a ring-shaped PSF with a radius of $\sim $$5$ arcsec, with the goal of reducing intrapixel variations and minimizing the impact of flat-field errors.

The data were gathered in staring mode, observing the target continuously without any dithering\footnote{This technique has been used for the same objective by several authors. See, for instance, Croll et al. (2010a, 2010b, 2011), de Mooij et al. (2011).}. Since OMEGA2000 has no auto-guider, using only the telescope tracking system, every time the xy-position on the detector drifted 3-4 pixels from the starting position, we performed a manual guiding correction in order to keep the target as much as possible at the same position on the detector. We acquired a series of data where every file has 15 individual images of 4s exposure each 
in order to increase the observing efficiency.

The staring mode observations were collected during approximately 5.4 continuous hours. Before and after this sequence, we also obtained focused images composed by five dither-point images each with the purpose of obtaining sky images for further subtraction.

The initial data reduction was performed using IRAF\footnote{IRAF is distributed by the National Optical Astronomy Observatory, operated by the Association of Universities for Research in Astronomy, Inc., under cooperative agreement with the National Science Foundation.} for the bad pixel removal, flat-fielding, and sky subtraction. We adopted a similar procedure for the sky subtraction to the one described by Croll et al. (2010a, 2010b, 2011), where we constructed a normalized sky map based on the previouly mentioned focused images. A single sky map was generated by normalizing the background of the stacked source-masked images and combining them by the median. From each image of the staring mode sequence, this sky map was scaled to the observed median background and then subtracted.

A circular aperture photometry was performed using the {\it aper.pro} procedure from the IDL Astronomy User's Library\footnote{IDL stands for Interactive Data Language - for further information, see http://www.ittvis.com/ProductServices/IDL.aspx; {\it aper.pro} is distributed by NASA - see http://idlastro.gsfc.nasa.gov/ for more details.}. We used a radius of 13 pixels to measure the stellar flux and the residual sky background was measured from a ring with radii of 20 and 37 pixels for the inner and outer annuli, for the target, and for all sufficiently bright stars in the field of view. Different apertures were tested, from 5 to 25 pixels, in steps of 0.5 pix, and we used the one that resulted in the optimal photometry, in order to obtain the magnitudes of the individual measurements, given by signal-to-noise ratio estimations. We also tested different sky annuli sizes; however, the final photometry did not present significant variations. 
Those stars selected as reference stars for the relative photometry are shown in 
Table \ref{refstars2MASS}. Stars presenting strong variations or any other odd behavior in their light curves were neglected.

\begin{table}
\caption{Reference stars used for the relative photometry.}\label{refstars2MASS}
\centering
\small
\begin{tabular}{lcc}
\hline\hline
Star No. & Identifier (2MASS) & K magnitude \\
\hline
1 & J23161168+3121302 & 9.743 \\
2 & J23152988+3124545 & 10.242 \\
3 & J23153263+3125204 & 11.277 \\
4 & J23154185+3125453 & 11.351 \\
5 & J23160820+3123526 & 11.311 \\
6 & J23161623+3125394 & 11.374 \\
7 & J23161550+3124310 & 10.182 \\
8 & J23162393+3126136 & 10.822 \\
9 & J23153719+3129053 & 11.059 \\
10 & J23155269+3127250 & 11.439 \\
11 & J23161392+3129131 & 10.611 \\
12 & J23162371+3128547 & 10.468 \\
13 & J23162211+3131385 & 8.295 \\
\hline
\end{tabular}
\end{table}

Since the reference stars have different magnitudes, we obtained the target's relative flux, $F$, by
   \begin{equation}\label{eqflux}
		F(t) = \dfrac {F_{tar}(t)} {\sum_{i=1}^{13}F_{ref,i}(t)} ,
   \end{equation}
where $F_{tar}(t)$ is the target's measured flux, and $F_{ref,i}(t)$ is the measured flux of reference star $i$ at a given time $t$. This relative flux was then normalized by its median value,
   \begin{equation}\label{eqflux2}
		f(t) = \dfrac {F(t)} {\tilde{F}} ,
   \end{equation}
where $\tilde{F}$ is the median value of target's relative flux and $f(t)$ is the target's normalized flux.

\section{Analysis}\label{analys}

\subsection{\bf Correction of systematic effects}\label{syseff}

\begin{figure}
   \centering
   \includegraphics[width=1.0 \columnwidth,angle=0]{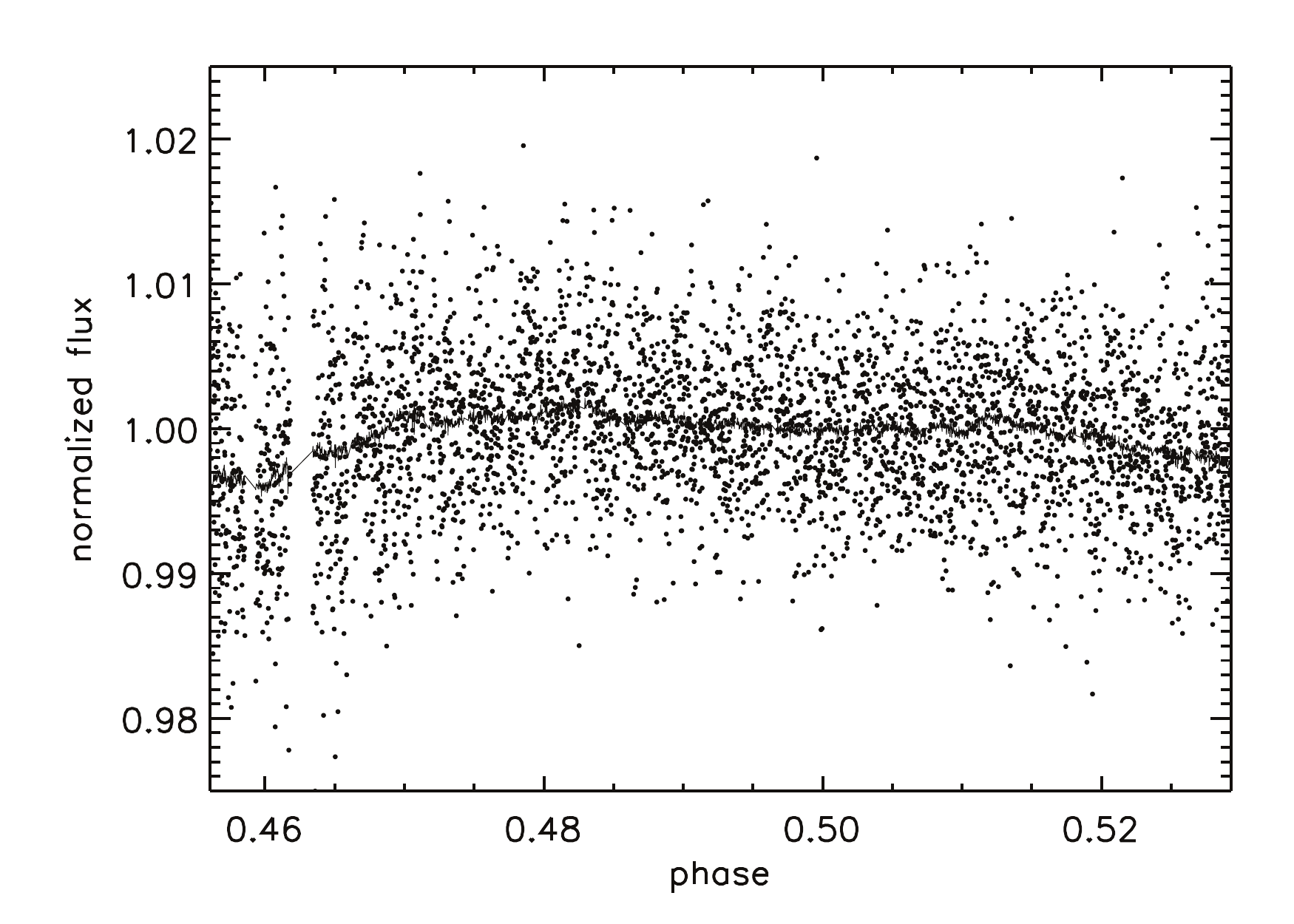}
   \includegraphics[width=1.0 \columnwidth,angle=0]{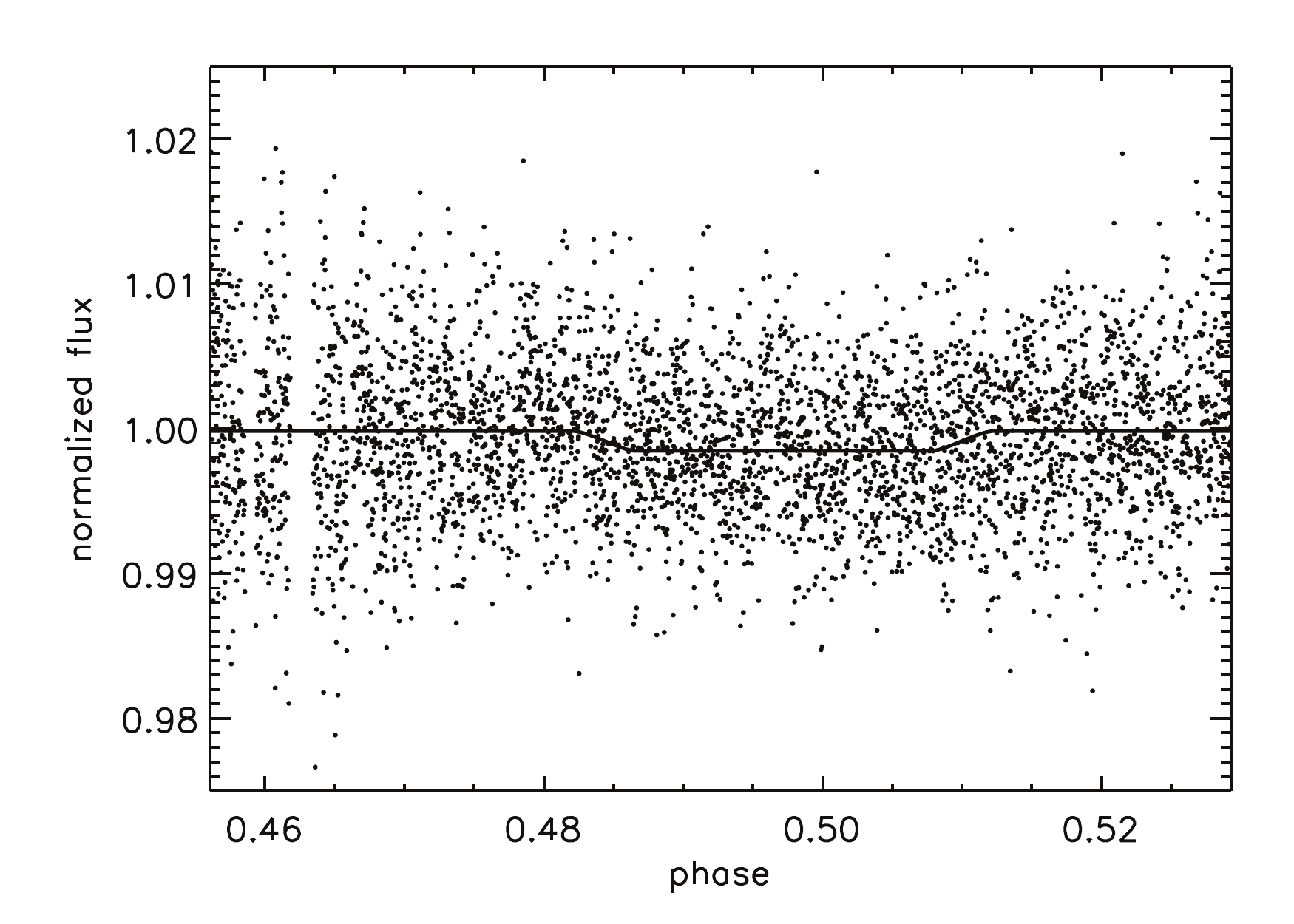}
      \caption{{\it Top panel:} Uncorrected relative light curve of WASP-10 in the Ks-band as a function of phase. {\it Bottom panel:} Light curve after correcting for systematic effects. The solid line shows the best fitting model.}
         \label{LCuncorrected}
   \end{figure}

The relative light curve of WASP-10 is shown in Figure \ref{LCuncorrected} (top panel), where the influence of systematic effects masks the eclipse signal. 
Before we can search for the secondary eclipse, we need to identify and remove significant systematic signals from the light curve. We searched for correlations between the stellar flux and several parameters, including XY-position of the centroid of WASP-10 on the detector, aperture correction\footnote{Since we used a fixed aperture of 13 pixels for the whole data set, an aperture correction was estimated by measuring the flux using a new radius around the target centered from a bondary of 4 sigma above the residual background. This was considered as a "defocused seeing", when multiplied by the plate scale.}, median background level measured before the sky map subtraction (see section \ref{obsred}), airmass and temperature and pressure of the detector.
%

In order to identify the parameters related to the visible trends in the observed light curve, we made use of a powerful statistical technique for dimensional evaluation in data sets called Principal Component Analysis (PCA). In a $R^{n}$ array, the PCA finds the linear combination (vector) of ${n}$ axes that best reproduces the data distribution in question. Following this technique (see, e.g., Morrison 1976), after constructing the variance-covariance and the correlation matrices, one obtains eigenvectors and eigenvalues that describe the whole data set, where the eigenvector with the highest eigenvalue is defined as the principal component, PCA1, which is the combination of parameters predominantly correlated. Subsequently, PCA2 has the second highest eigenvalue, showing a second pattern present in the data set in question, and so on. The first eigenvector (PCA1) is, hence, the vector that represents most of the variance in the data set, as a first approximation. This method was applied to our scientific case with the objective of minimizing the variance as much as possible with the fewest components.

As systematics are supposed to affect every image entirely, being present in all stars in the field-of-view, trends found in the light curve of any reference star should also be present in the light curve of WASP-10. This way, we can ensure that they are not intrinsic to our object. Therefore, we have calculated the PCAs of some of the reference stars listed in Table \ref{refstars2MASS}, selected by their 2MASS colors, minimizing differential refraction and other chromatic effects, and with no variability previously reported. These stars were treated individually, where their normalized fluxes were obtained from Eqs. \ref{eqflux} and \ref{eqflux2}, where ${F_{tar}}$ is now the measured flux of the reference star in question. These analyses have revealed significant correlations of the normalized flux with the star's y-position at the detector ($y_{c}$), aperture correction ($s$), airmass ($\sec z$), and background count level ($f_{bg}$).

We then calculated the PCAs for WASP-10 considering only the expected out-of-eclipse ($ooe$) part of the light curve, assuming a circular orbit, where the stellar flux is assumed to be constant. Similar correlations were found of the normalized flux with the same four parameters: $y_{c}$, $s$, $\sec z$, and $f_{bg}$. It is worth noting that only PCA1 was considered since the purpose here was to identify parameters that have a strong influence on the data and to eliminate only dominant patterns without compromising the collected signal.

We finally fitted for these systematics simultaneously by performing a multiple linear regression in IDL ({\it regress.pro}). This algorithm generated a polynomial of the form
   \begin{equation}\label{eqsys}
 		f_{ooe} = c_{0} + c_{1}y_{c,ooe} + c_{2}s_{ooe} + c_{3}\sec z_{ooe} + c_{4}f_{bg,ooe} ,
   \end{equation}
where $f_{ooe}$ is the out-of-eclipse flux of the target and $c_{k}$ are constants of the fit. 
This modeled trend was obtained considering only the out-of-eclipse portions of the light curve, preventing the eclipse signal from being removed with the systematics. Then, the model was applied to the in-eclipse portion and removed from the light curve.

The final detrended light curve is presented in Figure \ref{LCuncorrected} (bottom panel), where the data presents an improvement on the out-of-eclipse part of the light curve, going from a root-mean-square, RMS, per minute-integration of $2.7\times 10^{-3}$ to $2.2\times 10^{-3}$.

The periodogram of the detrended data was generated to look for small low-frequency variations in the light curve that could have remained, although nothing significant was found down to the binning frequency. Red noise was detected only in timescales greater than $\sim $9-10 minutes. We note that no significant periodicities were found.

  \begin{figure}
   \centering
   \includegraphics[width=1.0 \columnwidth,angle=0]{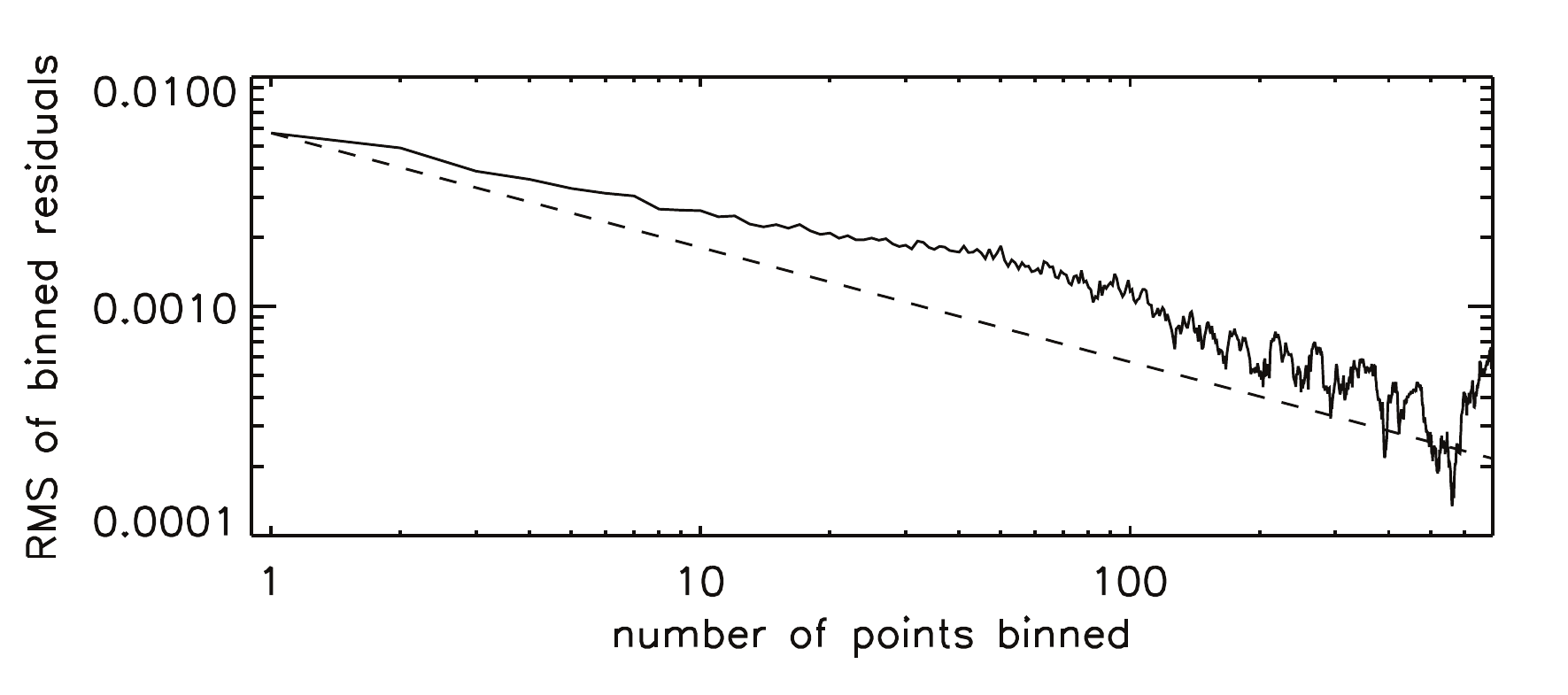}
      \caption{RMS of the residuals of the light curve for different bin sizes (see Section \ref{syseff}). The dashed line shows the limit expectation for normally distributed noise.}
         \label{rms}
   \end{figure}

In order to increase the precision of the photometry and reduce the error bars in the light curve, we have investigated the noise level (RMS) behavior, considering only the expected out-of-eclipse part, by binning the curve into different bin sizes from 1 to 700 points per bin. For the last case, the light curve would only have 3 points, since we have a total of 2249 out-of-eclipse individual measurements. The residuals behavior is shown in Figure \ref{rms}, along with the expected Gaussian noise given by one over the square root of the bin size. The photometry is still affected by other systematics, as the RMS of the residuals is slightly higher than the white noise limit. 
Considering that there is a significant contribution of red noise in the binned data, we have estimated the amount of additional noise that will be considered in the final uncertainties. 

Following the formalism presented by \cite{Pont06}, and discussed by \cite{CarterWinn09} as the ``time-averaging'' method, we find that the rescale factor, $\beta $, is given by
   \begin{equation}\label{beta}
 		\beta = \sqrt {1 + \left(\dfrac{\sigma_{r}}{\sigma_{w}} \right)^2} ,
   \end{equation}
where $\sigma_{r}$ and $\sigma_{w}$ are the red noise and the white noise, respectively.

As illustrated in Figure \ref{rms}, the RMS of the residuals decreases for greater bin sizes. We have binned the light curve considering the optimal RMS within a bin limit defined as half of the ingress duration. Thus, we performed our analysis binning the light curve with 127 points per bin ($\approx $8.47 minutes), corresponding to an interval of $\sim $$0.00190$ in phase, reaching a minimum RMS of $0.65\times 10^{-3}$. 
This RMS can be interpreted as the sum in quadrature of white noise and the red noise contribution of $0.51\times 10^{-3}$ and $0.41\times 10^{-3}$, respectively. This way, the rescale factor from Eq. \ref{beta} is of $\sim $$1.28$. 
However, if we estimate $\sigma_{r}$ and $\sigma_{w}$ using the system of equations dicussed by Winn et al. (2007, and references therein), which are
   \begin{equation}
 		\sigma_{1}^2 = \sigma_{w}^2 + \sigma_{r}^2 ,
   \end{equation}
   \begin{equation}
 		\sigma_{N}^2 = \frac{\sigma_{w}^2}{N} + \sigma_{r}^2 ,
   \end{equation}
where $\sigma_{1}$ and $\sigma_{N}$ are the standard deviation of the residuals and the standard deviation of the time-averaged residuals, respectively, then the $\beta $ value is much smaller, of around $\sim $$1.06$.

To avoid underestimating our errors, we consider the rescale factor calculated previously ($\sim $$1.28$), which represents an increase of $28\%$ in the uncertainties.


\subsection{\bf Modeling the secondary eclipse}\label{MAmodeling}

\begin{table}
\caption{Parameters of WASP-10b system considered in this work.}\label{SysParam}
\centering
\small
\begin{tabular}{lc}
\hline\hline
Parameter & Value \\
\hline
Normalized separation $a/R_{*}$	& 11.895$\pm $0.083 (1) \\
Planet-star radii ratio $R_{p}/R_{*}$	& 0.15758$^{+0.00036}_{-0.00039}$ (1) \\
Transit epoch $T_{0}$ (days)	& 2454664.038090$\pm $0.000048 (1) \\
Transit duration $T_{1-4}$ (hours)	& 2.2363$\pm $0.0051 (1) \\
Orbital period $P$ (days)	& 3.09272932$\pm $0.00000032 (1) \\
Orbital inclination $i$ (deg)	& 88.66$\pm $0.12 (1) \\
Semimajor axis $a$ (AU)	& 0.0375$\pm $0.0017 (1) \\
Stellar radius $R_{*}$ ($R_{\odot}$)	& 0.678$^{+0.028}_{-0.032}$ (1) \\
Planet radius $R_{p}$ ($R_{Jup}$)	& 1.039$^{+0.043}_{-0.049}$ (1) \\
Stellar $T_{\rm eff}$ (K)	& 4675$\pm $100 (2) \\
\hline
\end{tabular}
\tablefoot{References: (1) \cite{Barros13}; (2) \cite{Christian09}}
\end{table}

  \begin{figure}
   \centering
   \includegraphics[width=1.0 \columnwidth,angle=0]{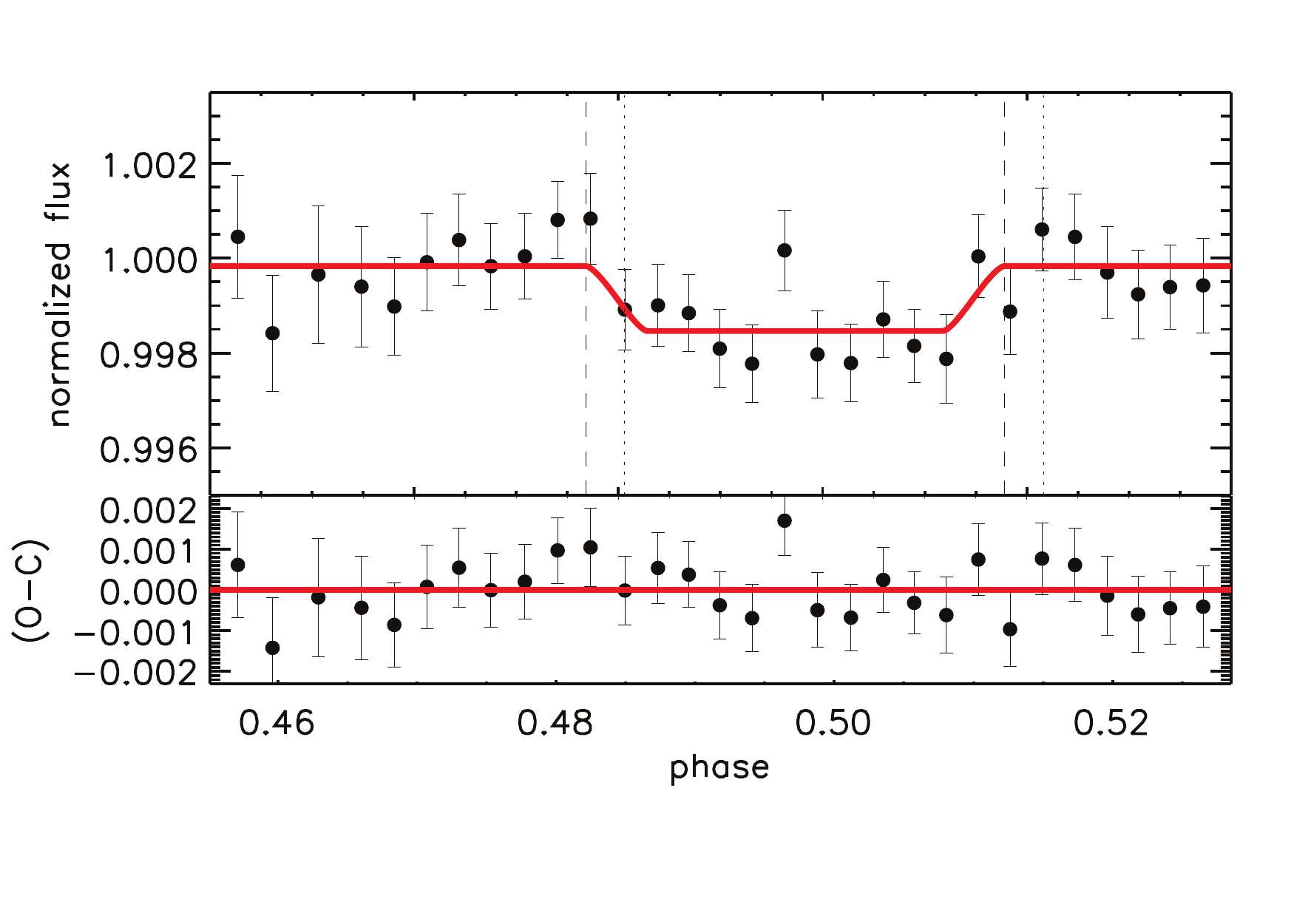}
      \caption{Secondary eclipse of WASP-10b in the Ks-band obtained in the MCMC analysis. The black circles show the data binned every $\sim $8.47 minutes (127 points per bin). The solid line represents the best model obtained from the MCMC analysis, resulting in an eclipse depth of $\Delta F = 0.137\%$, with a baseline level at $F_{bl}=0.99984$. The dotted lines illustrate the ingress and egress positions of the expected eclipse for a circular orbit, and the dashed lines show a phase shift of $\Delta \phi = -0.0028$, given by the best model. The residuals are presented in the lower panel.}
         \label{MAmodelFit}
   \end{figure}

  \begin{figure*}
   \centering
   \includegraphics[width=1.2 \columnwidth,angle=90]{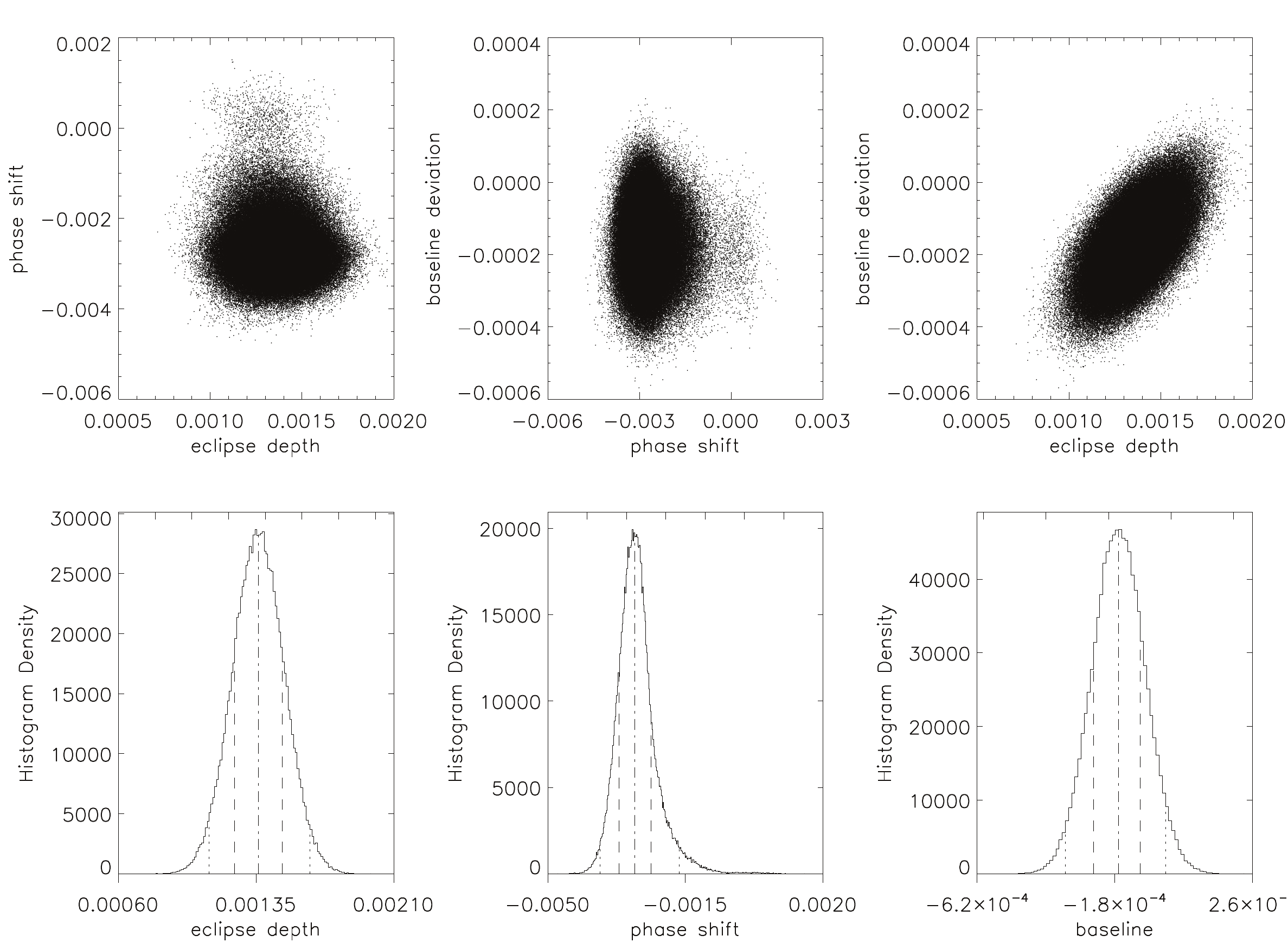}
      \caption{Distributions from the MCMC analysis, showing the correlation between parameters and the individual parameter histograms. The dashed and dotted lines show the error estimation for one and two sigma.}
         \label{mcmc_histo}
   \end{figure*}

The next step was to fit the binned light curve (with 127 points per bin) to obtain the secondary eclipse depth and detect a possible phase offset, due to an eccentricity. We used the occultation model from \cite{MandelAgol02}, assuming no limb-darkening. The parameters used were taken from \cite{Barros13} and are shown in Table \ref{SysParam}. We have performed two fitting procedures: grid models comparison and MCMC analysis. 
A third analysis was also performed, considering the unbinned light curve, unifying the detrend function and the occultation model in a joint fit, which will be presented later in this section.

The first analysis was performed by comparing the observations to a grid with more than $1.26\times 10^{7}$ models, generated in order to vary the expected phase of mid-eclipse $\phi_{c}$, the depth of the eclipse $\Delta F$, and the out-of-eclipse baseline level $F_{bl}$. This grid considers every configuration possible within a space of parameters that covers central phases from $0.47$ to $0.52$ in steps of $0.0001$, baseline levels from $0.998$ to $1.002$ in steps of $0.00005$, and eclipse depths from $-0.01\%$ to $0.3\%$ in steps of $0.001\%$, where the negative depths test the possibility of detecting a small increment in the measured flux instead of a decrement, which would be compatible with a non-detection. 
The best fit was defined as the model with the lowest $\chi^2$, presenting a depth of $\Delta F$ of $0.139\%$, a baseline level $F_{bl}$ of $0.99985$, and a phase shift $\Delta \phi $ of $-0.0028$. 
These results were used as inputs in the MCMC analysis.

For the second analysis, we fit for the same three parameters ($\Delta F$, $\Delta \phi $ and $F_{bl}$) using the MCMC method, generating four chains of $1.1\times 10^{6}$ each, with different initial conditions. 
To ensure that the initial conditions were not contaminating the final results, the first $1\times 10^{5}$ simulations from each chain were trimmed out of the analysis, remaining a total of $4\times 10^{6}$ when combining the results from all chains generated.

Figure \ref{MAmodelFit} presents the best model obtained from the MCMC analysis, with an eclipse depth $\Delta F$ of $0.137\%^{+0.010\%}_{-0.015\%}$, a phase offset $\Delta \phi $ of $-0.0028^{+0.0004}_{-0.0003}$ for a $1{\rm \sigma}$ detection, and a baseline level at $F_{bl}=0.99984^{+0.00006}_{-0.00008}$. Also shown are the ingress and egress positions expected for a circular orbit (dotted vertical lines) and the new ingress and egress given by the best model (dashed vertical lines). 
In order to have more reliable observational error bars, we have rescaled them such that the reduced $\chi^2$ was equal to unity by multiplying by factor of $\sqrt{(\chi^{2}_{dof})}$. In Figure \ref{mcmc_histo}, the correlation between the parameters obtained from the MCMC analysis are illustrated, along with the individual parameter histograms.

Taking into consideration the additional noise contribution discussed in Section \ref{syseff}, we have rescaled these uncertainties by the rescale factor presented in that section so as to have more reliable bars. Thus, we have obtained as results: $\Delta F = 0.137\%^{+0.013\%}_{-0.019\%}$, $\Delta \phi = -0.0028^{+0.0005}_{-0.0004}$, and $F_{bl} = 0.99984^{+0.00008}_{-0.00010}$.

  \begin{figure}
   \centering
   \includegraphics[width=0.90 \columnwidth,angle=0]{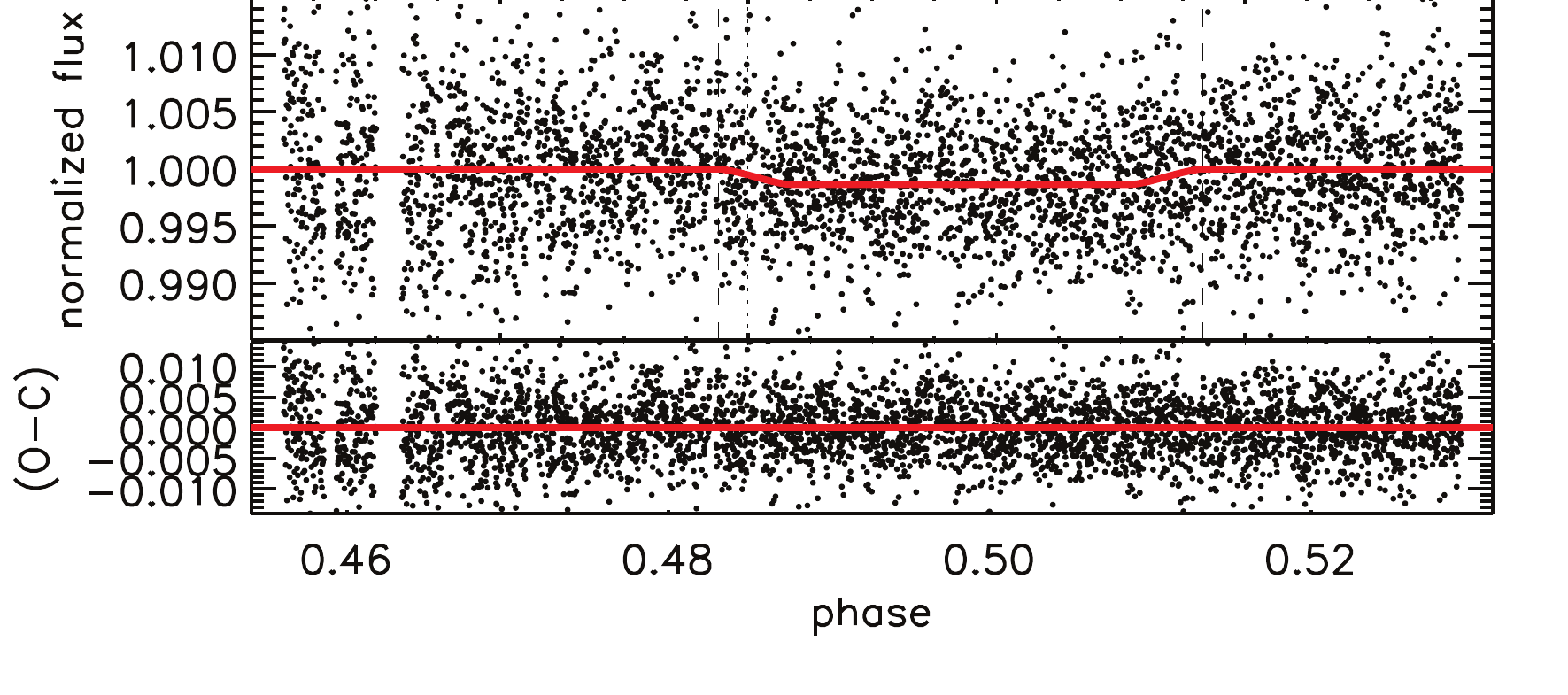}
   \includegraphics[width=0.90 \columnwidth,angle=0]{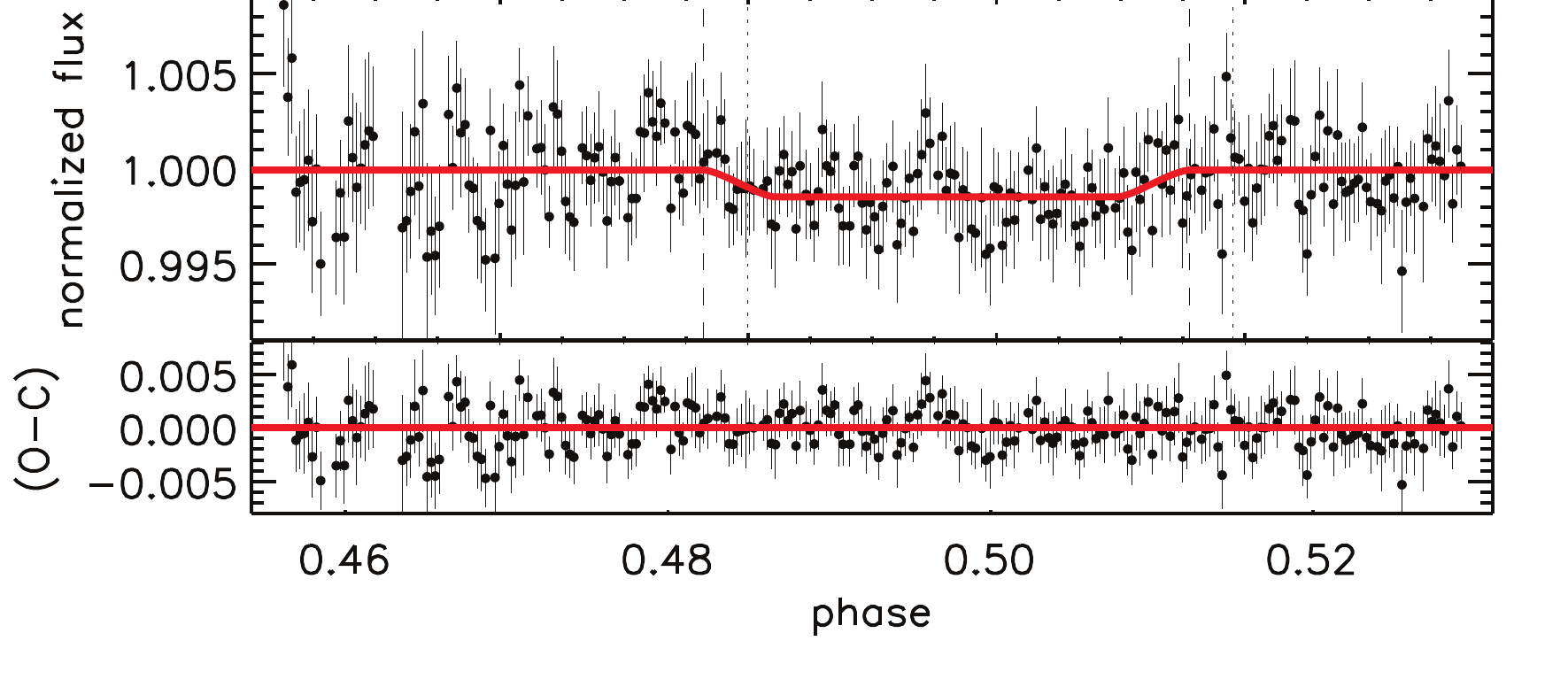}
      \caption{MCMC analyses for different bin sizes. From top to bottom, the black circles are the light curves with 1 (unbinned) and 15 points per bin (0.067 and 1 minute per bin, respectively). The solid lines show the model obtained for each case, showing the coherence of the results.}
         \label{MAmodelFitOther}
   \end{figure}

  \begin{figure}
   \centering
   \includegraphics[width=1.02 \columnwidth,angle=0]{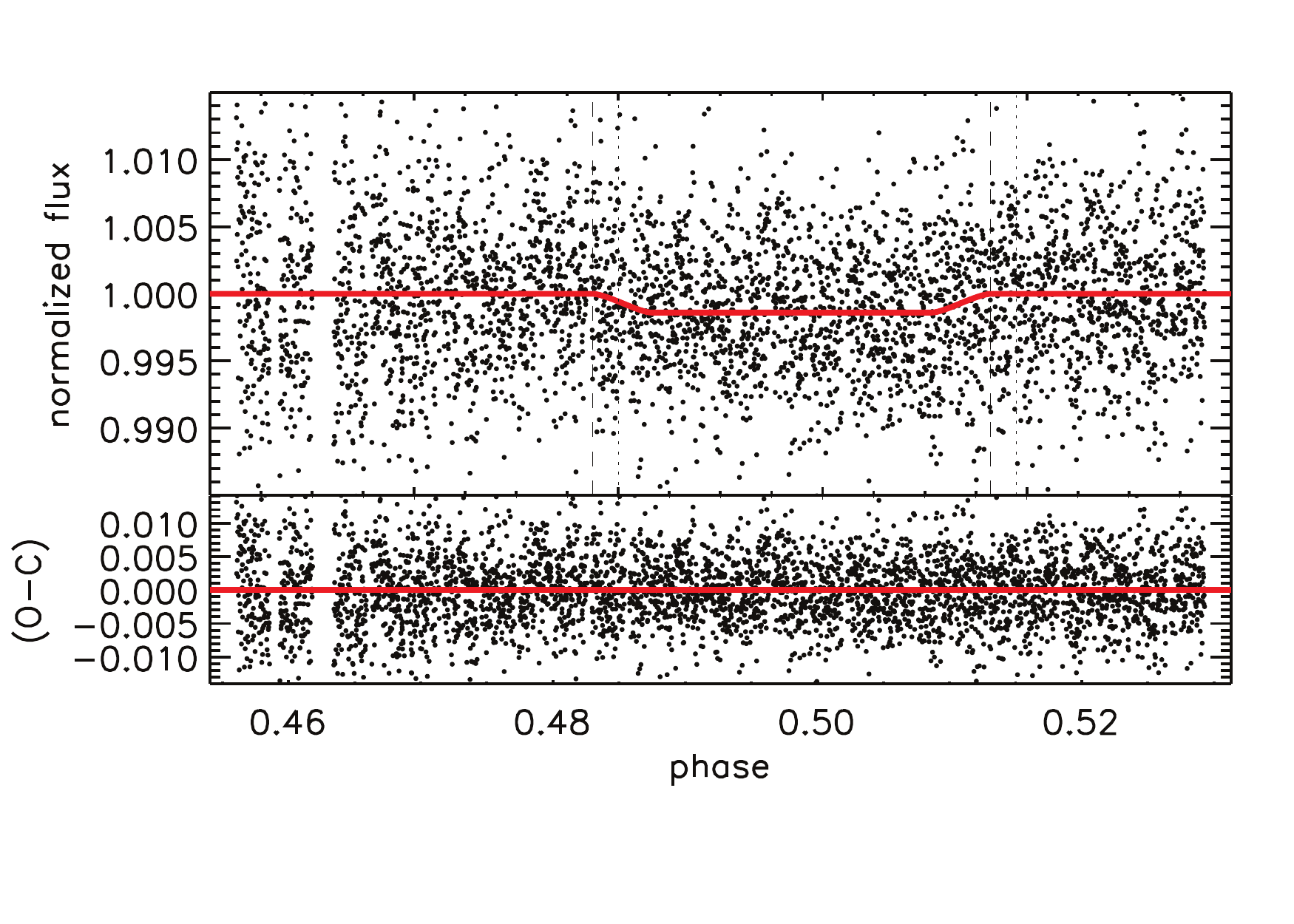}
      \caption{Secondary eclipse of WASP-10b in the Ks-band obtained in the joint-MCMC analysis. The small black circles show the unbinned data points. The solid line represents the best occultation model obtained from the joint-MCMC analysis, after removing the contribution of the systematics. The dotted lines illustrate the ingress and egress positions of the expected eclipse for a circular orbit, and the dashed lines show the phase shift given by the model. The residuals are presented in the lower panel.}
         \label{MAmodelFit2}
   \end{figure}

We also examined other cases to investigate the behavior of our results in function of the bin size considered. We applied the same MCMC analysis, first to the unbinned light curve with 3900 individual measurements, with a higher RMS ($5.7\times 10^{-3}$). The result provided a depth of $0.134\%^{+0.011\%}_{-0.013\%}$, a phase offset of $-0.0018^{+0.0017}_{-0.0003}$, and a baseline level at $F_{bl}=0.99998^{+0.00006}_{-0.00008}$ (for a $1{\rm \sigma}$ detection). 
Considering a bin of one minute (15 points per bin), we obtained $\Delta F = 0.140\%^{+0.018\%}_{-0.020\%}$, $\Delta \phi = -0.0027^{+0.0006}_{-0.0006}$, and $F_{bl}=0.99993^{+0.00009}_{-0.00012}$, with rescaled uncertainties.

In order to integrate the effect of the systematics into the uncertainties, we performed a joint fit of the unbinned light curve with the MCMC method as a third analysis (hereafter joint-MCMC). This joint-MCMC fits the eclipse and the systematics simultaneously, as
\begin{equation}
model=m_{occ} \times f_{sys} ,
\end{equation}
where $m_{occ}$ is the occultation model mentioned earlier in this section and $f_{sys}$ is similar to Eq. \ref{eqsys}, but now calculated for the whole data set.

In the same way as presented before, the joint-MCMC was done by running four chains of $1.1\times 10^{6}$ each. The initial conditions were defined as the results from the second analysis for the unbinned light curve, for the eclipse depth and the phase offset ($0.134\%$ and $-0.0018$, respectively). The baseline level was kept fixed ($F_{bl}=1.0$). As initial conditions for the $f_{sys}$ function, we have considered the contants obtained by the linear regression (see Sect. \ref{syseff}). The first $1\times 10^{5}$ simulations from each chain were also excluded from the analysis, as done in the previous MCMC.

This final analysis has provided the following results for the eclipse: $\Delta F = 0.139\%^{+0.011\%}_{-0.023\%}$ and $\Delta \phi = -0.0019^{+0.0019}_{-0.0002}$. Figure \ref{MAmodelFit2} shows the best occultation model obtained by this third analysis for the unbinned light curve, after removing the contribution of the systematics.

These analyses show that results from different fitting techniques and bin sizes are consistent and compatible with each other, within the error bars. Therefore, we have considered the results from the second analysis for binned light curve (with 127 points per bin).

%
%

  \begin{figure}
   \centering
   \includegraphics[width=1.02 \columnwidth,angle=0]{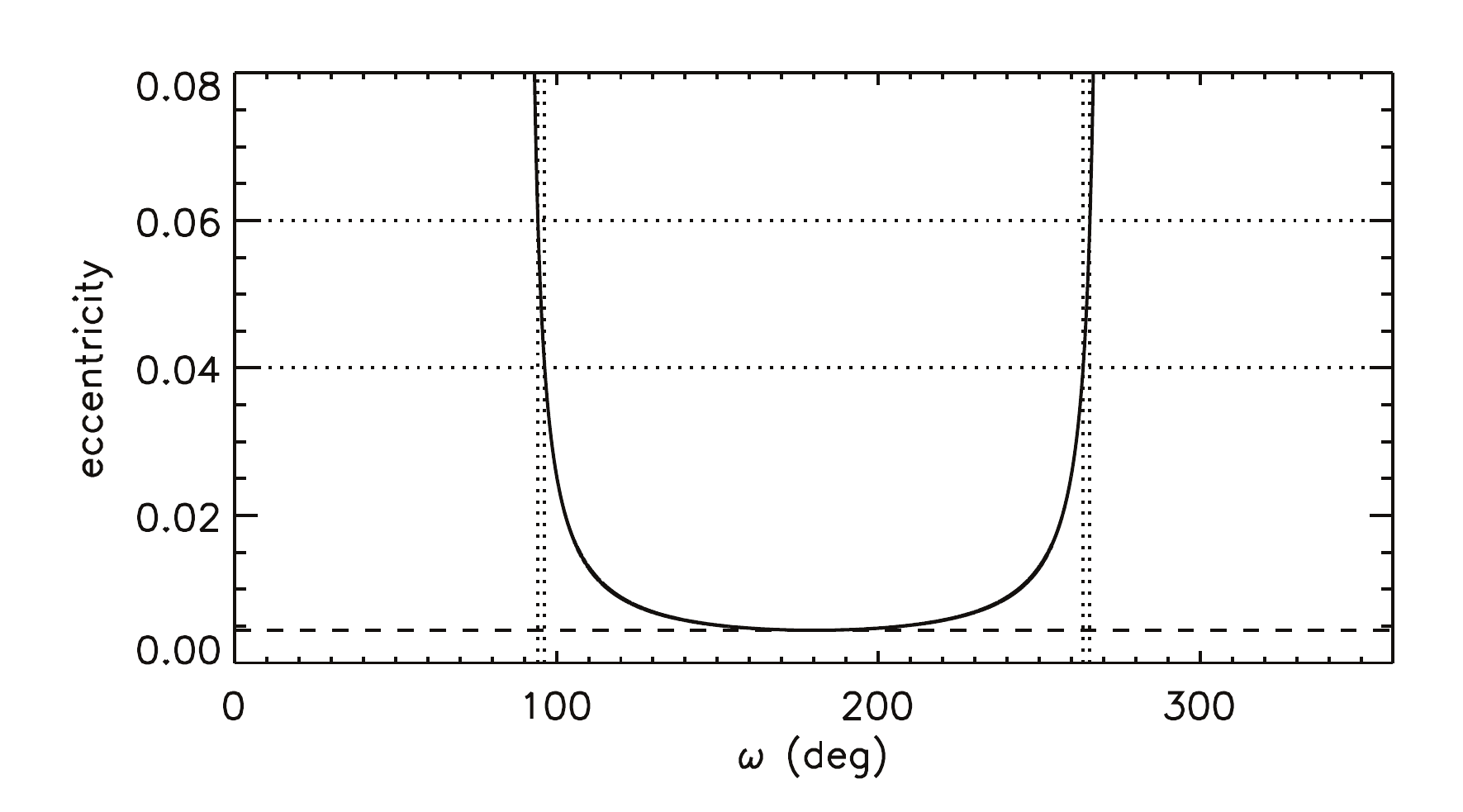}
      \caption{Testing different orbital configurations for $|e\cos{\omega}| \simeq 0.0044$. For very specific values of $\omega $, WASP-10b can have a small eccentricity, the order of $\sim 0.05$.}
         \label{ecc}
   \end{figure}

\section{Discussion}\label{discus}

\subsection{\bf A circular orbit for WASP-10b}

An observed phase difference between the center of the transit and the secondary eclipse ($\delta \phi =\phi_{ecl} - \phi_{tra}$) may suggest a non-zero eccentricity. 
The measured offset of $\Delta \phi =-0.0028$ means that the center of the secondary eclipse occured at phase $\phi_{ecl} =0.4972$ instead of at phase $\phi =0.5$, expected for a circular orbit. From the results, we were able to estimate the eccentricity of WASP-10b, using the following relation, in phase (Wallenquist 1950, L\'opez-Morales et al. 2010)
   \begin{equation}
		e\cos{\omega} =  \pi \cdot \dfrac {\delta \phi - 1/2} {1 + \csc^{2}(i)},
   \end{equation}
where $e$ is the eccentricity, $\omega $ the argument of periastron, and $i$ the orbital inclination. Considering $i$ from \cite{Barros13}, we obtain that $e\cos{\omega} \simeq -0.0044$. The negative sign indicates that $\omega $ is a value between $90$ and $270$ degrees. Considering the $\omega $ of $167.13^\circ $, published by \cite{Christian09} in WASP-10b's discovery paper, its orbit would have an eccentricity of $e \simeq 0.0045$. This value is fully consistent with a circular orbit, although it is in disagreement with their published eccentricity of $e \simeq 0.059$. Finally, if we use the value given by \cite{Johnson09} for $\omega $ of $153.3^\circ $, we get $e \simeq 0.0049$, also in favor of a null eccentricity and different from their result ($e \simeq 0.051$). These derived eccentricities are too small to be of any significance. 
Considering the phase offset found from our analysis, we cannot explain the eccentricities found in the literature.

The eclipse phase derived from our modeling should not be considered as evidence of an eccentric orbit. The phase offset found is small and there are systematic errors that were not taken into account in the formal phase uncertainty quoted. The offset is also slightly dependent on the analysis method, being frequently smaller than 0.0028.

Our observed phase suggests a circular orbit for WASP-10b, in agreement with previous results (Maciejewski et al. 2011a, Husnoo et al. 2012, Barros et al. 2013). 
Nevertheless, there are orbital configurations that can lead to the small eccentricities found in the literature for certain values of $\omega $. Figure \ref{ecc} illustrates the range of possibilities for the pair $e$ and $\omega $, according to our results for $e\cos{\omega}$, showing that in order to obtain an eccentricity of approximately $0.04$ to $0.06$ (Christian et al. 2009, Johnson et al. 2009, Husnoo et al. 2012), the argument of periastron should respect the intervals of $ 94.23^\circ \leq \omega \leq 96.35^\circ $ or $ 263.65^\circ \leq \omega \leq 265.77^\circ $. The values for $\omega $ published so far are placed outside the mentioned intervals and, thus we believe that the null eccentricity ($e=0$) represents an optimal solution.

\subsection{\bf Thermal emission of WASP-10b}

  \begin{figure}
   \centering
   \includegraphics[width=1.0 \columnwidth,angle=0]{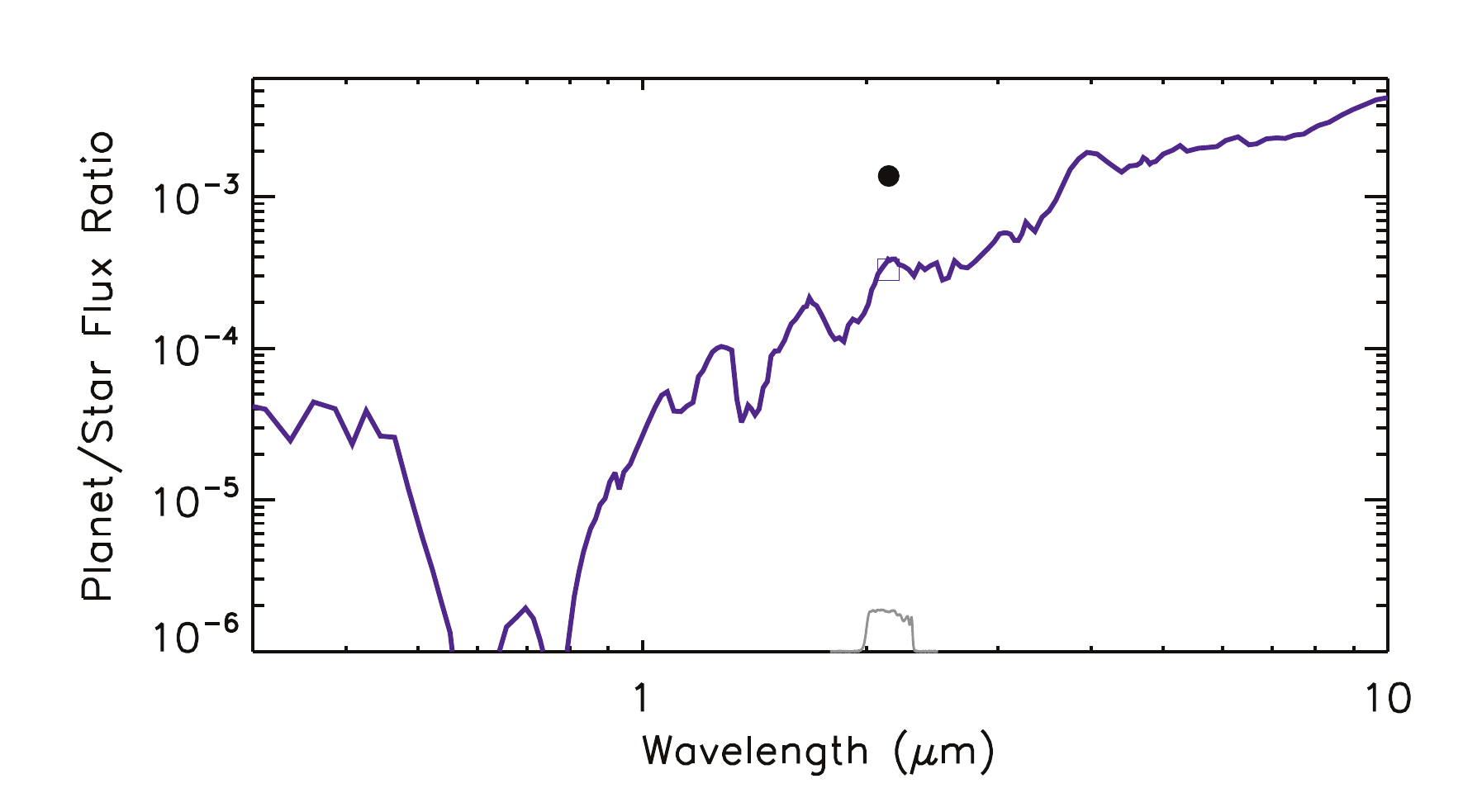}
      \caption{Model spectrum of thermal emission of WASP-10b with an instant reradiation over the dayside ($f=2/3$), computed without TiO/VO (solid line). The filled circle shows the measured planet-to-star flux ratio, in the Ks-band (at 2.14 $\mu $m). The error bars are inside the filled circle. The square represents the expected flux within the same band for the presented model. The Ks-band transmission curve (gray line) is shown at the bottom of the panel at arbitrary scale.}
         \label{spectra}
   \end{figure}

The MCMC analysis resulted in a planet-to-star flux ratio of $0.137\%$, as presented in section \ref{MAmodeling}. From that eclipse depth, assuming both components of the system, planet and star, emit as black-bodies, and considering the stellar parameters shown in Table \ref{SysParam}, we derived a Ks-band brightness temperature for WASP-10b of $T_{Ks}\simeq 1647^{+97}_{-131}$ K ($2\sigma$). The maximum expected equilibrium temperature\footnote{According to \cite{Lopez07}, the equilibrium temperature of an exoplanet is given by $T_{eq}=T_{s}({R_{s}}/{a})^{1/2}[f(1-A_{B})]^{1/4}$, where $T_{s}$ and $R_{s}$ are the stellar effective temperature and radius, $a$ is the orbital semimajor axis, $f$ is the reradiation factor, and $A_{B}$ is the planet's Bond albedo.} for the planet, assuming zero Bond albedo and instant reradiation ($A_{B}=0$ and $f=2/3$, respectively; see L\'opez-Morales \& Seager 2007) is $T_{eq}\simeq 1224$ K. This temperature is about $25\%$ cooler than the observed $T_{Ks}$. 

We compared the measured planet-to-star flux ratio with several atmospheric spectral models by Fortney et al. (2006, 2008), generated with different reradiation factors, computed without TiO/VO. However, none of the models used in this comparison was able to reproduce the observed high emission of WASP-10b. Figure \ref{spectra} shows, as an example, a model spectrum from Fortney and collaborators considering an instant reradiation over the dayside ($f=2/3$), computed without TiO/VO.

We have considered possible explanations for finding a higher temperature than the expected. 
The first one could be an overestimation of the eclipse depth. A recent study by \cite{Rogers13} has found that some of the light curve systematics found in ground-based data cannot always be successfully removed by de-correlation techniques and can introduce significant biases in the derived eclipse depth. These biases generally appear to be on the order of $\pm 10\%$. However, for the case of WASP-10b, even if the depth derived in Section \ref{MAmodeling} was overestimated by as much as $20\%$, the planet's observed temperature would only decrease by about $100$ K.

As mentioned before, several works have presented signatures of activity in WASP-10 (Smith et al. 2009, Maciejewski et al. 2011b, Barros et al. 2013). Hence, another explanation could be related to stellar activity. The intrinsic activity of the parent star could be affecting the planet's equilibrium temperature. It has been observed that activity can significantly affect the size of low-mass stars (e.g., L{\'o}pez-Morales 2007), resulting in a $10$-$15\%$ larger radius than the one predicted by models. However, even a $15\%$ increase in stellar radius would only increase the temperature of the planet by about $100$ K in this case. In addition, the activity-induced increase in radius is usually accompanied by a drop in the effective temperature of the star (Morales et al. 2010), so both effects compensate and the temperature of the planet will not vary significantly.

Nevertheless, since this is a young K-dwarf system with an age of $270 \pm 80$ Myr (Maciejewski et al. 2011a), strong flaring events are expected to occur more frequently and could affect the upper atmosphere of its planetary companion. As presented by \cite{Khodachenko07}, Coronal Mass Ejections (CMEs) may radically affect planetary environments due to their high speed, intrinsic magnetic fields, and increased plasma densities. The interaction of CMEs with upper planetary atmospheres can modify the thermospheric density and temperature structure, which can lead to atmospheric escape.

Lammer et al. (2006) suggested that Hot Jupiters with intrinsic magnetic moments of less than $10\%$ of Jupiter's magnetic moment cannot balance a dense CME plasma flow and can be strongly eroded by ion pick up and other non-thermal loss processes. Several studies have investigated such effects (Khodachenko et al. 2007, Lammer et al. 2007), concluding that constant UV flares and X-ray radiations are capable of heating and enhancing atmospheric escape, or even of destroying the exoplanet's atmosphere. 
An example of an exoplanet observed with an evaporating atmosphere is HD 189733b: a very hot Jupiter with semimajor axis of $0.0313$ AU and orbital period of $2.219$ days, orbiting an active K-dwarf (Bouchy et al. 2005). This system is very similar to the WASP-10b system ($a=0.0375$ AU, $P\simeq 3.093$ days), but a little older than WASP-10 (0.6 Gyr, Melo et al. 2006), and could help us to understand the present scenario. Recently, some models showed that most of the EUV and X-ray energy coming from an active central star makes the atmosphere escape the planetary gravitational potential (Sanz-Forcada et al. 2011, Owen \& Jackson 2012, Bourrier et al. 2013). Lecavelier Des Etangs et al. (2009, 2011) have found a low radio emission from HD 189733b, implying a weak planetary magnetic field, which could help to explain the evaporating atmosphere. Nevertheless, this is still a controversial subject and a deep study in the field would be necessary to check this hypothesis.

There is also the possibility of the planet actually being brighter than the expected equilibrium temperature. Several authors have published similar results, for instance, Rogers et al. (2009) on CoRoT-1b and C{\'a}ceres et al. (2011) on WASP-4b, among others. The most recent one is the work from \cite{Wang13} on WASP-43b, where they presented a Ks-band brightness temperature that was $\sim 300$ K higher than predicted. However, they have found a good agreement with other predictions from planetary atmospheric models.

\section{Conclusions}\label{concl}

In this work we investigated the hypothesis of a circular orbit for WASP-10b, by observing a secondary eclipse in the Ks-band. The data was acquired using the OMEGA2000 instrument at the 3.5-meter telescope at Calar Alto (Almer\'ia, Spain), in staring mode, with the telescope defocused. 
We made use of the Principal Component Analysis technique to identify systematic effects related to some visible trends in the photometry.

The final detrended light curve was fitted using a transit model from Mandel \& Agol (2002). Different analyses were performed, considering a grid of models and a MCMC analysis of several data binnings, which led to consistent results. The best model obtained from the MCMC analysis revealed an eclipse depth ($\Delta F$) of $0.137\%^{+0.013\%}_{-0.019\%}$, a phase offset ($\Delta \phi $) of $-0.0028^{+0.0005}_{-0.0004}$ for a $1{\rm \sigma}$ detection, with the baseline level ($F_{bl}$) at $0.99984^{+0.00008}_{-0.00010}$. The obtained phase offset leads to a value for $|e\cos{\omega}|$ of $0.0044$, which is ten times smaller than expected, and is fully consistent with a circular orbit.

Assuming the planet emits as a blackbody, from the measured planet-to-star flux ratio we derived the Ks-band brightness temperature for WASP-10b of $T_{Ks}\simeq 1647^{+97}_{-131}$ K, exceeding the maximum expected equilibrium temperature for the planet of $T_{eq}\simeq 1224$ K, considering zero Bond albedo and instant reradiation ($A_{B}=0$ and $f=2/3$, respectively). Several scenarios were considered in order to understand the high dayside thermal emission detected for WASP-10b. None of them, however, were conclusive.

\begin{acknowledgements}
PC would like to thank Dr. A. Gil de Paz, Dr. M. R. Zapatero Osorio and Dr. I. A. G. Snellen for all fruitful discussions and suggestions.

This research has been funded by Spanish grants AYA2010-21161-C02-02, AYA2012-38897-C02-01, and PRICIT-S2009/ESP-1496. PC, DB, JB and SH have received support from the RoPACS network during this research, a Marie Curie Initial Training Network funded by the European Commissions Seventh Framework Programme.

This article is based on data collected under Service Time program at the Calar Alto Observatory, the German-Spanish Astronomical Center, Calar Alto, jointly operated by the Max-Planck-Institut f\"ur Astronomie Heidelberg and the Instituto de Astrof\'isica de Andaluc\'ia (CSIC). We are very grateful to the CAHA staff for the super quality of the observations. 

This work has made use of the ALADIN interactive sky atlas and the SIMBAD database, operated at CDS, Strasbourg, France, and of NASA's Astrophysics Data System Bibliographic Services.

\end{acknowledgements}


\begin{thebibliography}{}

	\bibitem[Barros et al.(2013)]{Barros13} Barros, S. C. C., Bou{\'e}, G., Gibson, N. P.,
	  et al., 2013,
	  MNRAS, 760

	\bibitem[Bouchy et al.(2005)]{Bouchy05} Bouchy, F., Udry, S., Mayor, M.,
	  et al., 2005,
	  A\&A, 444, L15 

	\bibitem[Bourrier et al.(2013)]{Bourrier13} Bourrier, V., Lecavelier des Etangs, A., Dupuy, H.,
	  et al., 2013,
	  A\&A, 551, A63 

	\bibitem[C{\'a}ceres et al.(2011)]{Caceres11} C{\'a}ceres, C., Ivanov, V. D., Minniti, D.,
	  et al. 2011,
	  A\&A, 530, A5 

	\bibitem[Carter \& Winn (2009)]{CarterWinn09} Carter, J. A., \& Winn, J. N.,
	  2009,
	  ApJ, 704, 51 

	\bibitem[Christian et al.(2009)]{Christian09} Christian, D. J., Gibson, N. P., Simpson, E. K.,
	  et al., 2009,
      MNRAS, 392, 1585

	\bibitem[Croll et al.(2010a)]{Croll10a} Croll, B., Albert, L., Lafreniere, D.,
	  et al., 2010a,
	  ApJ, 717, 1084

	\bibitem[Croll et al.(2010b)]{Croll10b} Croll, B., Jayawardhana, R., Fortney, J. J.,
	  et al., 2010b,
	  ApJ, 718, 920

	\bibitem[Croll et al.(2011)]{Croll11} Croll, B., Lafreniere, D., Albert, L.,
	  et al., 2011,
	  AJ, 141, 30

	\bibitem[de Mooij et al.(2011)]{deMooij11} de Mooij, E. J. W., de Kok, R. J., Nefs, S. V.,
	  Snellen, I. A. G., 2011,
	  A\&A, 528, A49

	\bibitem[Dittmann et al.(2010)]{Dittmann10} Dittmann, J. A., Close, L. M., Scuderi, L. J.,
	  Morris, M. D., 2010,
	  ApJ, 717, 235

	\bibitem[Fortney et al.(2006)]{Fortney06} Fortney, J. J., Saumon, D., Marley, M. S.,
	  et al., 2006,
	  ApJ, 642, 495

	\bibitem[Fortney et al.(2008)]{Fortney08} Fortney, J. J., Lodders, K., Marley, M. S.,
	  et al., 2008,
	  ApJ, 678, 1419

	\bibitem[Husnoo et al.(2012)]{Husnoo12} Husnoo, N., Pont, F., Mazeh, T.,
	  et al., 2012,
	  MNRAS, 422, 3151

	\bibitem[Johnson et al.(2009)]{Johnson09} Johnson, J. A., Winn, J. N., Cabrera, N. E.,
	  Carter, J. A., 2009,
	  ApJL, 692, L100

	\bibitem[Khodachenko et al.(2007)]{Khodachenko07} Khodachenko, M. L., Ribas, I., Lammer, H.,
	  et al., 2007,
	  Astrobiology, 7, 167 

	\bibitem[Krej{\v c}ov{\'a} et al.(2010)]{Krejcova10} Krej{\v c}ov{\'a}, T., Budaj, J., 
	  \& Krushevska, V., 2010,
	  Contributions of the Astronomical Observatory Skalnate Pleso, 40, 77

	\bibitem[Lammer et al.(2006)]{Lammer06} Lammer, H., Khodachenko, M. L., Lichtenegger, H. I. M.,
	  et al., 2006,
	  European Planetary Science Congress, 388

	\bibitem[Lammer et al.(2007)]{Lammer07} Lammer, H., Lichtenegger, H. I. M., Kulikov, Y. N.,
	  et al., 2007,
	  Astrobiology, 7, 185 

	\bibitem[Lecavelier Des Etangs et al.(2009)]{Lecavelier09} Lecavelier Des Etangs, A.,
	  Sirothia, S. K., Gopal-Krishna, Zarka, P., 2009,
	  A\&A, 500, L51

	\bibitem[Lecavelier Des Etangs et al.(2011)]{Lecavelier11} Lecavelier Des Etangs, A.,
	  Sirothia, S. K., Gopal-Krishna, Zarka, P., 2011,
	  A\&A, 533, A50

	\bibitem[L{\'o}pez-Morales (2007)]{LopezMorales07} L{\'o}pez-Morales, M.,
	  2007,
	  ApJ, 660, 732 

	\bibitem[L{\'o}pez-Morales \& Seager (2007)]{Lopez07} L{\'o}pez-Morales, M., \& Seager, S.,	
	  2007,
	  ApJ, 667, L191 

	\bibitem[L{\'o}pez-Morales et al.(2010)]{Lopez10} L{\'o}pez-Morales, M., Coughlin, J. L., Sing, D. K.,
	  et al., 2010,
	  ApJ, 716, L36 

	\bibitem[Maciejewski et al.(2011a)]{Maciejewski11a} Maciejewski, G., Dimitrov, D., Neuh{\"a}user, R.,
	  et al., 2011a,
	  MNRAS, 411, 1204

	\bibitem[Maciejewski et al.(2011b)]{Maciejewski11b} Maciejewski, G., Raetz, S., Nettelmann, N.,
	  et al., 2011b,
	  A\&A, 535, A7

	\bibitem[Mandel \& Agol (2002)]{MandelAgol02} Mandel, K., Agol, E.,
	  2002,
	  ApJ, 580, L171

	\bibitem[Melo et al.(2006)]{Melo06} Melo, C., Santos, N. C., Pont, F., 
	  et al., 2006,
	  A\&A, 460, 251 

	\bibitem[Morales et al.(2010)]{Morales10} Morales, J. C., Gallardo, J., Ribas, I.,
	  et al., 2010,
	  ApJ, 718, 502 

	\bibitem[Morrison (1976)]{Morrison76} Morrison, D. F., 1976, 
	  Multivariate Statistical Methods. McGraw-Hill Book Co., Singapore

	\bibitem[Owen \& Jackson (2012)]{Owen2012} Owen, J. E., \& Jackson, A. P.,
	  2012,
	  MNRAS, 425, 2931 

	\bibitem[Pont et al.(2006)]{Pont06} Pont, F., Zucker, S., \& Queloz, D.,
	  2006,
	  MNRAS, 373, 231 

	\bibitem[Rogers et al.(2009)]{Rogers09} Rogers, J. C., Apai, D., L{\'o}pez-Morales, M., Sing, D. K.,
	  et al., 2009,
	  ApJ, 707, 1707 

	\bibitem[Rogers et al.(2013)]{Rogers13} Rogers, J., L{\'o}pez-Morales, M., Apai, D., \& Adams, E.,
	  2013,
	  ApJ, 767, 64 

	\bibitem[Sanz-Forcada et al.(2011)]{Sanz11} Sanz-Forcada, J., Micela, G., Ribas, I.,
	  et al., 2011,
	  A\&A, 532, A6

	\bibitem[Smith et al.(2009)]{Smith09} Smith, A. M. S., Hebb, 
	  L., Collier Cameron, A., et al., 2009,
	  MNRAS, 398, 1827

	\bibitem[Wallenquist (1950)]{Wallenquist1950} Wallenquist, {\AA} ., 
	  1950,
	  Ark. Astron., 1, 59

	\bibitem[Wang et al.(2013)]{Wang13} Wang, W., van Boekel, R., Madhusudhan, N., 
	  et al., 2013,
	  ApJ, 770, 70 

	\bibitem[Winn et al.(2007)]{Winn07} Winn, J. N., Holman, M. J., Henry, G. W., 
	  et al., 2007,
	  AJ, 133, 1828 


\end{thebibliography}
\end{document}